\documentclass{article}

\usepackage{arxiv}

\usepackage[utf8]{inputenc} 
\usepackage[T1]{fontenc}    
\usepackage{tipa} 
\usepackage{hyperref}       
\usepackage{url}            
\usepackage{booktabs}       
\usepackage{amsfonts}       
\usepackage{nicefrac}       
\usepackage{microtype}      
\usepackage{graphicx}
\usepackage{natbib}
\usepackage{doi}
\usepackage{amsmath}        
\usepackage{enumitem}       
\usepackage{cleveref}       
\usepackage{booktabs}       
\usepackage{makecell}       
\usepackage{soul}           

\usepackage{multirow} 

\usepackage{tikz}
\usetikzlibrary{shapes.geometric, arrows.meta, positioning, calc}

\title{SwiftF0: Fast and Accurate Monophonic Pitch Detection}

\newif\ifuniqueAffiliation
\uniqueAffiliationtrue

\ifuniqueAffiliation 
\author{ \href{https://orcid.org/0000-0002-9821-1636}{\includegraphics[scale=0.06]{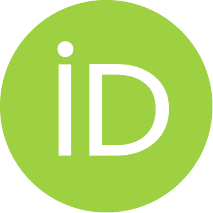}\hspace{1mm}Lars~Nieradzik}\\
	\texttt{l.nieradzik@gmail.com} \\
}
\else
\usepackage{authblk}

\newbox{\orcid}\sbox{\orcid}{\includegraphics[scale=0.06]{orcid.pdf}} 
\author[1]{%
	\href{https://orcid.org/0000-0000-0000-0000}{\usebox{\orcid}\hspace{1mm}David S.~Hippocampus\thanks{\texttt{hippo@cs.cranberry-lemon.edu}}}%
}
\author[1,2]{%
	\href{https://orcid.org/0000-0000-0000-0000}{\usebox{\orcid}\hspace{1mm}Elias D.~Striatum\thanks{\texttt{stariate@ee.mount-sheikh.edu}}}%
}
\affil[1]{Department of Computer Science, Cranberry-Lemon University, Pittsburgh, PA 15213}
\affil[2]{Department of Electrical Engineering, Mount-Sheikh University, Santa Narimana, Levand}
\fi

\renewcommand{\shorttitle}{SwiftF0}

\hypersetup{
pdftitle={SwiftF0: Fast and Accurate Monophonic Pitch Detection},
pdfsubject={eess.AS, cs.LG},
pdfauthor={Lars~Nieradzik},
pdfkeywords={Pitch Detection, Neural Networks, STFT},
}

\begin{document}
\maketitle

\begin{abstract}
Accurate and real-time monophonic pitch estimation in noisy conditions, particularly on resource-constrained devices, remains an open challenge in audio processing. We present \emph{SwiftF0}, a novel, lightweight neural model that sets a new state-of-the-art for monophonic pitch estimation. Through training on diverse speech, music, and synthetic datasets with extensive data augmentation, SwiftF0 achieves robust generalization across acoustic domains while maintaining computational efficiency. SwiftF0 achieves a 91.80\% harmonic mean (HM) at 10 dB SNR, outperforming baselines like CREPE by over 12 percentage points and degrading by only 2.3 points from clean audio. SwiftF0 requires only 95,842 parameters and runs approximately 42x faster than CREPE on CPU, making it ideal for efficient, real-time deployment. To address the critical lack of perfectly accurate ground truth pitch in speech corpora (which typically rely on algorithmic estimators or laryngograph signals), we introduce \emph{SpeechSynth}. This synthetic speech dataset, generated by a phoneme-level TTS model, provides exact, on-demand ground-truth pitch curves, enabling more robust model training and evaluation. Furthermore, we propose a unified metric, combining six complementary performance measures for comprehensive and reliable pitch evaluation, and release an open-source pitch benchmark suite. A live demo of SwiftF0 is available at \url{https://swift-f0.github.io/}, the source code at \url{https://github.com/lars76/swift-f0}, and the benchmark framework at \url{https://github.com/lars76/pitch-benchmark}.
\end{abstract}

\keywords{Monophonic Pitch Detection \and Neural Networks \and STFT \and F0 Estimation}

\section{Introduction}

Monophonic pitch estimation is a fundamental task in audio signal processing, with applications spanning speech analysis, music information retrieval, and audio synthesis. Despite decades of research, accurate and efficient pitch estimation in noisy environments remains challenging.

In recent years, deep learning-based approaches such as CREPE \citep{kim2018crepeconvolutionalrepresentationpitch} have significantly advanced the state-of-the-art in pitch tracking. However, these models often come with considerable computational costs, making them less suitable for real-time or resource-limited applications. Moreover, their performance can degrade under adverse acoustic conditions, limiting their robustness in practical scenarios.

In this work, we introduce SwiftF0, a lightweight and highly accurate algorithm for monophonic pitch estimation. SwiftF0 is designed to be both computationally efficient and robust to noise, while maintaining or surpassing the pitch estimation accuracy of state-of-the-art models.

Our main contributions are as follows:

\begin{enumerate}
    \item \textbf{SwiftF0: A novel, highly efficient, and robust neural model} for monophonic pitch estimation that outperforms existing state-of-the-art methods like CREPE in accuracy, especially under noisy conditions, while being significantly faster and more parameter-efficient.
    \item \textbf{SpeechSynth: A novel synthetic speech dataset} providing exact ground truth pitch annotations, addressing the limitations of existing speech datasets which often rely on estimated or approximate pitch labels.
    \item \textbf{A unified evaluation framework} encompassing a new harmonic mean (HM) metric combining six complementary performance aspects, along with a comprehensive open-source benchmark suite for rigorous and reproducible evaluation of pitch estimators.
\end{enumerate}

\begin{figure}[htbp]
    \centering
    \includegraphics[width=\textwidth]{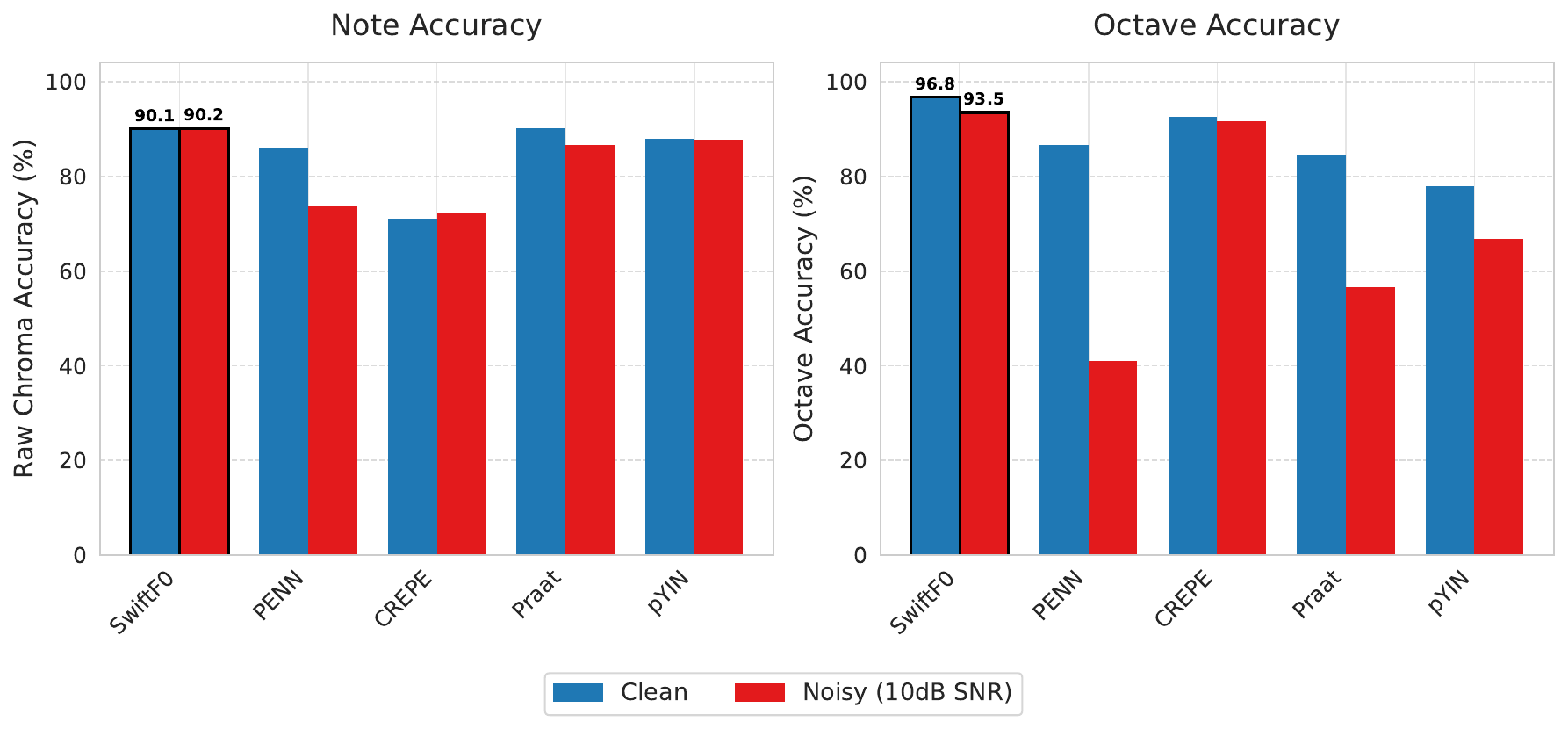}
    \caption{SwiftF0 is a new monophonic pitch estimation algorithm achieving superior pitch accuracy compared to current methods. It performs robustly in noise (10dB SNR), surpassing CREPE, yet uses significantly fewer resources: 95,842 parameters and ~42x faster processing (vs. CREPE's 22M parameters). Our evaluation spans diverse held-out datasets including Vocadito, Bach10-mf0-synth, and our novel SpeechSynth dataset.}
    \label{fig:teaser}
\end{figure}

As shown in Figure \ref{fig:teaser}, SwiftF0 demonstrates superior performance and efficiency across diverse evaluation datasets.

Our results suggest that SwiftF0 is a compelling alternative for real-time and embedded pitch estimation applications, enabling high-quality pitch tracking even in resource-constrained settings.

\section{Related Work}

Existing approaches to monophonic pitch estimation fall into two main categories: classical signal processing methods and modern deep learning-based techniques. We briefly review prominent algorithms in both groups.

\subsection{Signal Processing}

Traditional F0 estimation methods exploit the periodic nature of harmonic signals through mathematical analysis. Autocorrelation-based approaches like Praat~\citep{BoersmaWeenink2025} and pYIN/YIN~\citep{deCheveigne2002yin, Mauch2014} detect periodicity in the time domain, while RAPT~\citep{Talkin1995} uses normalized cross-correlation for robust tracking. Frequency-domain methods such as SWIPE~\citep{Camacho2007} match input spectra to sawtooth templates, achieving high accuracy on clean signals. Hybrid approaches like YAAPT~\citep{Zahorian2008} combine time and frequency domain information through dynamic programming. While computationally efficient and interpretable, these methods typically struggle with noise, complex timbres, and parameter sensitivity, motivating the shift toward data-driven approaches.

\subsection{Neural Network}

Deep learning has revolutionized pitch estimation by learning robust representations directly from data. CREPE~\citep{kim2018crepeconvolutionalrepresentationpitch} pioneered this paradigm using a CNN that processes 1024-sample raw audio excerpts to predict pitch probability distributions over 360 bins spanning six octaves. Despite achieving state-of-the-art accuracy (>90\% at 10-cent tolerance) and strong noise robustness on certain datasets, CREPE's computational demands (22M parameters) limit real-time deployment.

Recent work has focused on improving efficiency while maintaining accuracy. PENN~\citep{morrison2023cross} optimizes CREPE's architecture through finer binning (1440 bins at 5-cent resolution), categorical cross-entropy loss, and entropy-based periodicity estimation, achieving better real-time CPU performance. BasicPitch~\citep{bittner2022lightweightinstrumentagnosticmodelpolyphonic} targets automatic music transcription with a lightweight architecture (16.8K parameters) using Constant-Q Transform inputs, though it focuses on polyphonic note transcription rather than monophonic pitch tracking.

However, existing neural approaches have still limitations in practical deployment scenarios. Most methods do not prioritize real-time performance constraints. Additionally, many approaches are not designed to handle the broad spectrum of audio signals encountered in real-world applications, potentially limiting their generalization across diverse acoustic conditions. Furthermore, existing methods process either raw waveforms or full spectral representations without exploiting the efficiency gains possible through strategic frequency selection for the specific task of pitch estimation.

\textbf{Our Approach.} SwiftF0 addresses these limitations through strategic frequency band selection (46.875-2093.75 Hz), removing 74\% of spectral bins for efficient STFT processing. Our compact 95.8K parameter architecture directly optimizes continuous pitch estimation through joint classification and regression training. By training across diverse speech, music, and synthetic datasets with extensive data augmentation, SwiftF0 achieves robust generalization while maintaining real-time efficiency.

\section{Method}

\subsection{Architecture Overview}

Our pitch detection system employs a convolutional neural network that processes spectral representations of audio signals. The architecture comprises three main components: (1) Short-Time Fourier Transform (STFT) preprocessing that converts audio to time-frequency representations, (2) a stack of 2D convolutional layers that extract hierarchical features from the spectrogram, and (3) a frequency projection layer that maps spectral features to pitch bin predictions.

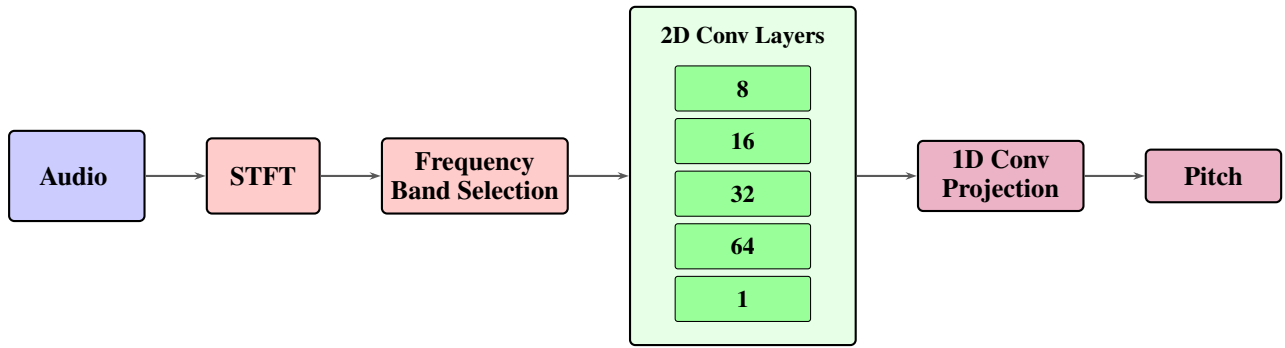
\begin{figure}[htbp]
  \centering
  \begin{tikzpicture}[
    node distance=0.8cm,
    every node/.style={font=\small},
    blockBlue/.style={
      rectangle, draw=black, thick, rounded corners=2pt,
      minimum width=2cm, minimum height=1cm, align=center,
      fill=blue!20, font=\bfseries
    },
    blockRed/.style={
      rectangle, draw=black, thick, rounded corners=2pt,
      minimum width=2cm, minimum height=1cm, align=center,
      fill=red!20, font=\bfseries
    },
    blockGreen/.style={
      draw=black, thick, rounded corners=2pt,
      fill=green!10, minimum width=3cm, minimum height=4.5cm
    },
    filterBox/.style={
      rectangle, draw=black, fill=green!40, rounded corners=1pt,
      minimum width=1.8cm, minimum height=0.6cm, align=center,
      font=\footnotesize\bfseries
    },
    blockPurple/.style={
      rectangle, draw=black, thick, rounded corners=2pt,
      minimum width=1.8cm, minimum height=0.7cm, align=center,
      fill=purple!30, font=\bfseries
    },
    arr/.style={
      -{Stealth[length=1.5mm,width=1mm]}, thick, draw=gray!70!black
    },
    wave/.style={thick, blue!70!black}
  ]
    \node (audio) [blockBlue, minimum width=1.8cm, minimum height=1.2cm, inner sep=2pt] {
      Audio
    };
    
    \node (stft) [blockRed, right=of audio, minimum width=1.5cm] {STFT};
    
    \node (band) [blockRed, right=of stft, minimum width=2.2cm] {Frequency\\Band Selection};
    
    \node (conv) [blockGreen, right=of band] {};
    \node[anchor=north,font=\bfseries\small]
      at ($(conv.north)+(0,-0.15cm)$) {2D Conv Layers};
    
    \foreach \i [count=\n from 0] in {8,16,32,64,1} {
      \node[filterBox] at
        ($ (conv.north)+(0,-1.1cm-\n*0.7cm)$) {\i};
    }
    
    \node (proj) [blockPurple, right=of conv, minimum width=2.2cm] {1D Conv\\Projection};
    
    \node (pitch) [blockPurple, right=0.8cm of proj.east] {Pitch};
    
    \draw[arr] (audio.east) -- (stft.west);
    \draw[arr] (stft.east) -- (band.west);
    \draw[arr] (band.east) -- (conv.west);
    \draw[arr] (conv.east) -- (proj.west);
    \draw[arr] (proj.east) -- (pitch.west);
    
  \end{tikzpicture}
  \caption{Pitch detection model architecture showing data flow from audio input through STFT, frequency band selection, 2D convolutional layers (with feature map counts), 1D convolutional projection, and pitch output. The model contains 95,842 parameters, which is significantly fewer than models like CREPE \citep{kim2018crepeconvolutionalrepresentationpitch}, with approximately 22 million parameters.}
  \label{fig:pitch_model_arch}
\end{figure}

Figure~\ref{fig:pitch_model_arch} illustrates the complete network architecture, showing the data flow from raw audio input through spectral preprocessing and convolutional feature extraction to the final pitch output.

\subsection{Input Representation and Preprocessing}

We use Short-Time Fourier Transform (STFT) magnitude spectrograms as input rather than raw waveform data. This choice is motivated by pitch detection being fundamentally a frequency estimation task, where spectral analysis provides direct access to harmonic content. Furthermore, many papers \citep{bittner2022lightweightinstrumentagnosticmodelpolyphonic,6854950,ren2022fastspeech2fasthighquality,luo2021lightspeechlightweightfasttext} have shown that frequency representations outperform neural networks operating on raw data. The STFT representation also reduces computational complexity while providing a structured time-frequency input that facilitates model convergence.

Given a discrete-time audio signal $x[n]$, we compute the STFT as:

$$\text{STFT}(x)[m, k] = \sum_{n=0}^{N-1} x[n + mH] w[n] e^{-j2\pi kn/N}$$

where $w[n]$ is a Hann window of length $N = 1024$, $H = 256$ is the hop size, $m$ indexes time frames, $k$ indexes frequency bins, and the sampling rate is $f_s = 16{,}000$ Hz. These parameters provide a time resolution of $\nicefrac{256}{16000} = 16$ ms and frequency resolution of $\nicefrac{16000}{1024} = 15.625$ Hz. We experimented with various parameter configurations and found these values to offer the best balance between computational efficiency and model performance.

The magnitude spectrogram is obtained as $X[m, k] = |\text{STFT}(x)[m, k]|$, retaining only bins \( k \in \{0, \dots, N/2\} \) corresponding to positive frequencies. We focus analysis on a frequency range relevant to human vocal pitch and most musical instruments:

$$f_{\text{min}} = 46.875~\text{Hz}, \quad f_{\text{max}} = 2093.75~\text{Hz}$$

These bounds are mapped to discrete STFT bin indices:

$$k_{\text{min}} = \text{round}\left(\frac{f_{\text{min}} \cdot N}{f_s}\right), \quad k_{\text{max}} = \text{round}\left(\frac{f_{\text{max}} \cdot N}{f_s}\right)$$

yielding $k_{\text{min}} = 3$ and $k_{\text{max}} = 134$. We slice the spectrogram to retain only these bins, producing a reduced representation $\tilde{X} \in \mathbb{R}^{T \times K}$, where $T$ is the number of time frames and $K = k_{\text{max}} - k_{\text{min}} + 1 = 132$. This slicing removes 74\% of the $513$ frequency bins, which significantly reduces memory usage as well as inference time.

Logarithmic compression is applied to align with human auditory perception and handle the wide dynamic range of audio signals:

$$S[m, k] = \log(\tilde{X}[m, k] + \epsilon)\,,$$

where $\epsilon = 10^{-8}$ prevents numerical issues with zero-valued bins.

\subsection{Network Architecture}

The spectrogram $S \in \mathbb{R}^{T \times 132}$ is processed through a convolutional neural network consisting of five 2D layers with feature map dimensions $[8, 16, 32, 64, 1]$. Each layer is followed by batch normalization and ReLU activation. This structure enables the network to learn hierarchical representations, from local spectral patterns in early layers to complex harmonic structures in deeper ones.

All convolutional layers use $5 \times 5$ kernels with stride 1 and "same" padding, preserving temporal and frequency dimensions throughout the network. Stacking five such layers results in a receptive field of $21 \times 21$ bins at the final layer. Given the frequency resolution of approximately 15.6 Hz per bin, this receptive field covers about $21 \times 15.6 \approx 328\, \text{Hz}$ along the frequency axis, sufficient to capture the fundamental frequency and several harmonics. The final layer projects the features to a single channel, producing an STFT-shaped output of size $T \times 132$ that emphasizes pitch-relevant information.

This output undergoes frequency projection via a 1D convolutional layer along the frequency axis, mapping the $K = 132$ linearly-spaced STFT frequency bins to $B = 200$ logarithmically-spaced pitch bins. Although logarithmic compression was applied to magnitude values, the underlying STFT frequency bins remain linearly spaced in Hz. This projection transforms the linear frequency representation to a perceptually-motivated logarithmic pitch scale suitable for classification.

The pitch bin centers $f_b$ are distributed logarithmically between the frequency bounds:

$$f_b = f_{\text{min}} \cdot 2^{b \cdot \Delta}, \quad \Delta = \frac{\log_2(f_{\text{max}} / f_{\text{min}})}{B - 1}$$

where $b \in [0, B-1]$. Since each bin advances by a fixed ratio in frequency, this corresponds to a constant resolution in perceived pitch. The effective pitch resolution per bin can be computed by converting the step size $\Delta$ to cents using the relation $1200 \cdot \Delta$, yielding approximately $1200\cdot\frac{\log_2(2093.75/46.875)}{200-1}
\approx 33.1$ cents per bin across the full range (1200 cents = 1 octave).

In contrast, CREPE \citep{kim2018crepeconvolutionalrepresentationpitch} defines its 360 pitch bins directly on a cent scale. Each bin corresponds to a pitch value of

\[
f_b = 32.7 \cdot 2^{\,b\,\Delta_c},
\quad
\Delta_c = \frac{20}{1200},
\quad
b \in \{0,\dots,359\},
\]

resulting in 360 bins spaced at uniform 20-cent intervals. This covers a range from approximately 32.7 Hz (C1) to 1975.5 Hz (B7), spanning six octaves.

While CREPE's fixed 20-cent binning provides a finer discrete resolution than our 33.1 cents per bin, the ultimate pitch precision in such models often relies more on post-processing techniques (e.g., local averaging) that refine predictions beyond the discrete bin spacing. Empirically, we found that increasing our model's bin count beyond 200 yielded no improvement in final pitch estimation accuracy.

\subsection{Training Objective}

The model is trained using a loss function that combines discrete pitch classification with continuous frequency regression. While classification encourages the model to assign high probability to the correct pitch bin, it limits predictions to fixed frequency steps, resulting in quantization errors. For example, the model might produce a sharp peak with probability 1.0 at a single bin. Adding the regression loss encourages the model to distribute probability mass to neighboring bins, such as 0.9 for one bin and 0.1 for an adjacent bin. This results in distributions that reflect pitch uncertainty more accurately and improve estimation at inference time when the full distribution is used rather than just the argmax.

The discrete component employs categorical cross-entropy loss over pitch bin indices:

$$\mathcal{L}_{\text{CE}} = -\frac{1}{T} \sum_{m=1}^{T} \sum_{b=1}^{B} y_{m,b} \log \hat{p}_{m,b}\,,$$

where $y_{m,b}$ is the one-hot encoded target for frame $m$ and bin $b$, and $\hat{p}_{m,b} = \text{softmax}(Z[m, b])$ represents the predicted probability distribution over pitch bins.

The continuous component applies L1 loss in log-frequency space to encourage precise frequency estimation:

$$\mathcal{L}_{\text{cents}} = \frac{1}{T} \sum_{m=1}^{T} \left| \hat{f}_{\log}[m] - \log(f_{\text{true}}[m]) \right|\,,$$

where the expected log-frequency is computed as:

$$\hat{f}_{\log}[m] = \sum_{b=1}^{B} \hat{p}_{m,b} \cdot \log(f_b)\,.$$

This L1 loss directly optimizes cents error up to a constant factor. The expected log-frequency $\hat{f}_{\log}[m]$ provides a differentiable estimate computed from softmax probabilities, encouraging the model to concentrate probability mass around the correct frequency.

The total training loss combines both components as:

$$\mathcal{L}_{\text{total}} = \mathcal{L}_{\text{CE}} + \lambda \mathcal{L}_{\text{cents}}\,.$$

We set $\lambda = 1$, giving equal weighting to classification and regression losses. Empirically, varying $\lambda$ around this value has minimal impact on model performance.

\subsection{Inference}

At inference time, the model outputs, for each frame $m$, a score vector $Z[m, b]$ over pitch bins. Applying a softmax yields the predicted distribution $\hat{p}_{m,b} = \mathrm{softmax}(Z[m, b])$, which serves as the basis for both the pitch estimate and a confidence score.

To convert this distribution into a continuous frequency estimate within $[f_{\text{min}}, f_{\text{max}}]$, we adopt the local expected value method from CREPE \citep{kim2018crepeconvolutionalrepresentationpitch}. This approach first identifies the most probable pitch bin,

$$
b^* = \arg\max_b Z[m, b],
$$

and then defines a local window centered on this peak:

$$
\mathcal{W}_m = \{ b : |b - b^*| \leq w \},
$$

where $w = 9$ defines the half-width of the window.

Within this window, we normalize the softmax probabilities to obtain a local distribution:

$$
\tilde{p}_{m,b} = \frac{\hat{p}_{m,b}}{\sum_{b' \in \mathcal{W}_m} \hat{p}_{m,b'}}.
$$

We then compute the continuous pitch estimate as a weighted average of the bin center frequencies:

$$
\hat{f}[m] = \sum_{b \in \mathcal{W}_m} \tilde{p}_{m,b} \cdot f_b.
$$

Furthermore, we derive a voicing confidence score based on how much total probability mass lies within the local window:

$$
c[m] = \sum_{b \in \mathcal{W}_m} \hat{p}_{m,b}.
$$

This confidence score reflects how concentrated the distribution is around the peak and is used for voicing decisions. In practice, we apply a threshold of approximately 90\%.

Because the continuous estimate $\hat{f}[m]$ is computed using the same expected log-frequency formulation as in the regression loss $\mathcal{L}_{\text{cents}}$, our training objective directly optimizes the local expected value employed at inference. Consequently, alternative decoding methods like the Viterbi algorithm did not yield improvements over the local expected value in our experiments.

\subsection{Datasets}

To ensure robust generalization across diverse audio domains, we selected a range of datasets that collectively cover speech, instrumental, synthetic, and real-world conditions. A key challenge in pitch detection is the absence of consistent, high-quality ground truth across most datasets. Each dataset we evaluated introduces trade-offs in terms of accuracy, realism, and domain coverage.

\begin{itemize}
\item \textbf{NSynth \citep{nsynth2017}:} Contains single-note synthetic audio from musical instruments with accurate pitch labels. While useful for controlled experiments, it lacks temporal and spectral complexity typical of real-world acoustic environments.

\item \textbf{PTDB-TUG \citep{556d2ea3b9c94141a6e0e785972cde62}:} A speech dataset providing recordings with laryngograph signals that directly capture vocal fold vibrations. However, the laryngograph device does not supply pitch annotations. Instead, ground truth pitch is derived by high-pass filtering the raw laryngograph signals to remove low-frequency artifacts followed by pitch extraction using the RAPT algorithm. Both filtering and RAPT processing can introduce errors. In our experiments, we excluded approximately 347 files (7.4\% of the dataset) with noticeable artifacts.

\item \textbf{MIR-1k \citep{5153305}:} Features vocal excerpts with pitch contours initially extracted algorithmically (e.g., via YIN), followed by manual correction. Despite the correction, labels still reflect algorithmic biases, limiting their value for learning true pitch from raw audio.

\item \textbf{MDB-STEM-Synth \citep{salamon2017f0}:} Offers musically structured synthetic audio with accurate pitch annotations. Though valuable for pitch learning, its synthetic nature may not fully capture real-world acoustic variability.

\item \textbf{Vocadito \citep{bittner2021vocaditodatasetsolovocals}:} Employs pitch annotations derived from pYIN, refined through manual verification. Like MIR-1k, its reliance on algorithmic estimates imposes limitations on the purity of its ground truth.

\item \textbf{Bach10-mf0-synth \citep{salamon2017f0}:} Similar in nature to MDB-STEM-Synth, it offers high-quality pitch labels for synthesized musical performances but lacks the complexity of real audio environments.
\end{itemize}

While each dataset has its own strengths, none alone provides complete coverage of real-world pitch variation. A notable gap is the absence of a synthetic speech dataset with perfectly accurate ground truth pitch. To address this, we introduce a new dataset, \textbf{SpeechSynth}.

SpeechSynth was constructed using a LightSpeech model \citep{luo2021lightspeechlightweightfasttext} trained on a combination of AISHELL-3 \citep{shi2021aishell3multispeakermandarintts} (85.62 hours) and Biaobei \citep{biaobei_dataset_2017} (11.86 hours), totaling 97.48 hours of Mandarin speech from diverse speakers. We specifically chose Mandarin due to its tonal nature, which presents a challenging and comprehensive test case for accurate pitch estimation.

Unlike many standard text-to-speech (TTS) models trained at the word level, our system is trained at the phonetic level, using a set of 54 phones. This inventory includes a wide variety of phonemes, such as affricates, nasal finals, retroflexes, and rhotic vowels, alongside their associated tones.

The LightSpeech model adopts a lightweight convolutional architecture that offers performance comparable to transformer-based models such as FastSpeech2 \citep{ren2022fastspeech2fasthighquality}, while maintaining lower computational overhead. With full control over phonetic sequences and speaker identity, SpeechSynth allows the generation of arbitrary speech samples with exact pitch labels, making it an effectively infinite dataset for training and evaluation.

For model training and hyperparameter optimization, we employed 5-fold group cross-validation across NSynth, PTDB-TUG, MIR-1k, MDB-STEM-Synth, and SpeechSynth. Groups were defined by unique speakers, instruments, or musical pieces to prevent data leakage between training and validation folds.

For evaluation, we used three held-out datasets: Vocadito \citep{bittner2021vocaditodatasetsolovocals}, Bach10-mf0-synth \citep{salamon2017f0}, and an independently generated SpeechSynth test set. The latter contains audio samples not seen during training. This setup ensures a rigorous assessment of the model’s generalization capability across both synthetic and natural domains.

\subsection{Data Augmentation}

Our data augmentation pipeline generates robust training examples by introducing controlled noise and varying gain levels, while focusing exclusively on voiced frames, since only these contain reliable pitch information. Unvoiced frames are handled during inference by discarding any low‑confidence pitch predictions. For each example, we extract a random 0.5‑second audio segment centered on a voiced frame and first adjust its amplitude by sampling a gain $G$ uniformly from $[-6,+6]$ dB, yielding

$$
x_{\text{scaled}}(t) = x(t)\cdot 10^{G/20}.
$$

We then create realistic background interference by blending environmental recordings from the CHiME‑Home dataset \citep{7336899} with synthetic white Gaussian noise $n_{\text{gauss}}(t)\sim\mathcal{N}(0,1)$. A mixing coefficient $\alpha$ is drawn at random from $[0,1]$, and the noise is formed as

$$
n_{\text{bg}}(t) = \sqrt{\alpha}\,n_{\text{env}}(t) + \sqrt{1-\alpha}\,n_{\text{gauss}}(t).
$$

To achieve natural‑sounding signal‑to‑noise ratios, a target SNR is selected uniformly from $[10,30]$ dB, and the corresponding noise‑scaling factor $\gamma$ is computed. The final augmented waveform is obtained by adding the appropriately scaled noise to the gain‑adjusted signal and clamping the result to $[-1,+1]$ to prevent clipping:

$$
x_{\text{aug}}(t) = \mathrm{clamp}\bigl(x_{\text{scaled}}(t) + \gamma\,n_{\text{bg}}(t), -1, +1\bigr).
$$

By varying loudness, background noise and SNR, our model is exposed to a diverse set of conditions during training, which improves its ability to generalize to new acoustic environments.

\subsection{Metric}

Traditional metrics for pitch estimation, such as Raw Pitch Accuracy (RPA), Raw Chroma Accuracy (RCA), and voicing F1-score, each measure different aspects of performance. However, none of them provide a complete picture of a pitch detector's overall capability. For example, RPA checks how close the estimated pitch is to the correct one, but it treats all large mistakes the same, whether the error is a subtle tritone or an entire octave. RCA is even more lenient: it only checks if the note name (pitch class) is correct, for instance, any C note regardless of its octave. Because of this, RCA does not penalize octave errors at all. This highlights a significant gap: no single traditional metric fully captures the performance of a pitch detector.

To address this limitation and provide a single performance indicator, we define the harmonic mean (HM) of six complementary components:

\begin{equation*}
    \label{eq:HM}
    \mathrm{HM} = \frac{6}{\displaystyle\sum_{i=1}^{6} \frac{1}{c_i}},
    \qquad
    c_i\in\{\mathrm{RPA},\mathrm{CA},\mathrm{P},\mathrm{R},\mathrm{OA},\mathrm{GEA}\}.
\end{equation*}
\begin{enumerate}[label=\textbf{\arabic*.}, wide=0pt, leftmargin=*]
\item \textbf{Raw Pitch Accuracy (RPA).}
    The fraction of voiced frames whose estimate lies within 50 cents of the ground truth:
    \begin{equation*}
        \mathrm{RPA} = \frac{\#\{\;|\Delta_t| < 50\;\}}{N},
        \qquad
        \Delta_t = 1200\log_2\!\left(\frac{f_{\mathrm{pred},t}}{f_{\mathrm{true},t}}\right)\,,
    \end{equation*}
    where N = total \# of voiced frames. RPA gives the percentage of voiced frames for which the predicted pitch is considered "correct enough" because its error is very small (less than 50 cents).
\item \textbf{Cents Accuracy (CA).}
    A fine-grained precision measure that penalizes larger deviations exponentially:
    \begin{equation*}
        \mathrm{CA} = \exp\!\left(-\frac{\bar{\Delta}}{500}\right),
        \qquad
        \bar{\Delta} = \frac{1}{N}\sum_{t=1}^{N}|\Delta_t|.
    \end{equation*}
    This captures subtle pitch deviations. The coefficient 500 was chosen to prevent HM from dropping to zero on challenging datasets.
\item \textbf{Voicing Precision (P).}
    The fraction of predicted voiced frames that are truly voiced:
    \begin{equation*}
        \mathrm{P} = \frac{\mathrm{TP}}{\mathrm{TP}+\mathrm{FP}},
    \end{equation*}
    where TP and FP denote true positives and false positives, respectively. This measures errors, where an algorithm incorrectly labels unvoiced frames as voiced.
\item \textbf{Voicing Recall (R).}
    The fraction of truly voiced frames that are detected:
    \begin{equation*}
        \mathrm{R} = \frac{\mathrm{TP}}{\mathrm{TP}+\mathrm{FN}},
    \end{equation*}
    where FN denotes false negatives. This captures errors, where the system fails to detect actual pitched frames.
\item \textbf{Octave Accuracy (OA).}
    Robustness against octave errors, defined as relative errors exceeding 40\,\% or absolute deviations of 1100–1300 cents:
    \begin{equation*}
        \mathrm{OA} = \exp\!\left(-10\cdot\frac{\#\{\;t:\text{octave error at }t\;\}}{N}\right).
    \end{equation*}
    This targets octave confusions, where the estimated pitch class is correct but placed in the wrong octave. As OA is particularly important, we use a high coefficient of $10$.
\item \textbf{Gross Error Accuracy (GEA).}
    Overall tracking stability, penalizing deviations larger than 200 cents:
    \begin{equation*}
        \mathrm{GEA} = \exp\!\left(-5\cdot\frac{\#\{\;|\Delta_t| \ge 200\;\}}{N}\right).
    \end{equation*}
    This covers all other large pitch deviations exceeding two semitones. Since gross errors are less critical than octave errors, we use a smaller decay factor of 5.
\end{enumerate}

A pitch algorithm must perform well on all six metrics to achieve a high score. Excelling in five while failing in one yields a value near zero. As additional information, we also report RCA, but it is not included in the harmonic mean for two reasons. First, because RPA demands both correct pitch class and correct octave, any frame counted by RPA will necessarily be counted by RCA. Hence,
$\text{RPA} \leq \text{RCA}$. Second, once RPA is already high, RCA contributes almost no additional information to the harmonic mean.

\section{Evaluation}

This section presents a comprehensive evaluation of pitch detection algorithms across three datasets: Bach10-mf0-synth, Vocadito \citep{bittner2021vocaditodatasetsolovocals}, and SpeechSynth. The evaluation encompasses both neural network-based and traditional signal processing approaches under two distinct testing conditions: clean audio and noisy environments. To simulate challenging acoustic conditions, we introduce additive noise at 10 dB SNR using samples from the CHiME-Home dataset \citep{7336899}, which contains realistic domestic background noise scenarios. All results reported represent the average performance across the three datasets.

We evaluate a diverse set of pitch detection algorithms, including classical signal processing techniques (Praat, RAPT, SWIPE, YAAPT, and pYIN) and modern deep learning-based models (CREPE, PENN, and BasicPitch). These algorithms were selected based on their popularity and the availability of reliable open-source implementations. For consistency and reproducibility, we use standardized libraries. Notably, we use Parselmouth \citep{Jadoul2018} for Praat, SPTK \citep{SPTK2025} for RAPT and SWIPE, librosa \citep{BMcFee2023_librosa} for pYIN, pYAAPT \citep{pYAAPT2023} for YAAPT, and TorchCREPE \citep{TorchCREPE2023} for the PyTorch-based implementation of CREPE. Our proposed method, SwiftF0, is included in the comparison as a competing approach.

\subsection{Performance Under Noisy Conditions}

Table~\ref{tab:averaged_results_noise} presents the comparative performance of all evaluated algorithms under noisy conditions (10 dB SNR).

\begin{table}[htbp]
    \centering
    \resizebox{\textwidth}{!}{%
    \begin{tabular}{llccccccc}
        \toprule
        \textbf{Category} & \textbf{Algorithm} & \makecell{\textbf{F1 Score} $\uparrow$} & \makecell{\textbf{CA} $\uparrow$} & \makecell{\textbf{RPA} $\uparrow$} & \makecell{\textbf{RCA} $\uparrow$} & \makecell{\textbf{OA} $\uparrow$} & \makecell{\textbf{GEA} $\uparrow$} & \makecell{\textbf{Harmonic}\\\textbf{Mean} $\uparrow$} \\
        \midrule
        \multirow{4}{*}{\makecell{Neural\\Networks}}
        & PENN \citep{morrison2023cross}          & 82.23 & 66.63 & 72.17 & 73.90 & 40.95 & 47.16 & 59.87 \\
        & CREPE \citep{kim2018crepeconvolutionalrepresentationpitch}    & 75.13 & 88.56 & 72.27 & 72.43 & 91.66 & 81.57 & 78.97 \\
        & BasicPitch \citep{bittner2022lightweightinstrumentagnosticmodelpolyphonic}    & 79.30 & 51.92 & 22.87 & 24.00 & 31.57 & 9.40  & 29.47 \\
        & SwiftF0 (Ours)       & 89.53 & \textbf{93.90} & \textbf{89.90} & \textbf{90.17} & \textbf{93.52} & \textbf{95.45} & \textbf{91.80} \\
        \midrule
        \multirow{5}{*}{\makecell{Signal\\Processing}}
        & Praat \citep{BoersmaWeenink2025}         & 89.47 & 80.72 & 83.57 & 86.73 & 56.55 & 71.42 & 76.10 \\
        & RAPT \citep{Talkin1995}          & 91.80 & 66.80 & 76.03 & 78.40 & 60.65 & 72.23 & 74.70 \\
        & SWIPE \citep{Camacho2007}         & 87.73 & 76.87 & 81.47 & 85.13 & 47.71 & 66.92 & 71.43 \\
        & YAAPT \citep{Zahorian2008}         & \textbf{92.57} & 72.14 & 77.30 & 85.03 & 36.54 & 57.30 & 65.30 \\
        & pYIN \citep{Mauch2014}          & 80.13 & 84.56 & 86.97 & 87.77 & 66.83 & 79.04 & 79.99 \\
        \bottomrule
    \end{tabular}%
    }
    \caption{Averaged performance under noisy conditions (10 dB SNR). SwiftF0 achieves the highest scores in 6 of 7 metrics ($\uparrow$ = higher is better; best in \textbf{bold}). The "Harmonic Mean" is a 6-component aggregate (RPA, CA, P, R, OA, GEA). Precision (P) and Recall (R) are used in its calculation  but are not explicitly shown for brevity.}
    \label{tab:averaged_results_noise}
\end{table}

SwiftF0 demonstrates exceptional performance across multiple evaluation metrics. The algorithm achieves the best performance in six out of seven metrics. The harmonic mean of 91.80\% represents an improvement of around 12\% to the next-best performing algorithm.

Among traditional signal processing methods, YAAPT achieves the highest F1 score (92.57\%) but suffers from lower Cents Accuracy (72.14\%) and gross error accuracy (57.30\%). CREPE demonstrates competitive performance with relatively high cents accuracy (88.56\%) and octave accuracy (91.66\%). BasicPitch exhibits poor performance across all metrics, as it was developed for polyphonic music transcription and not monophonic pitch estimation.

\subsection{Performance Under Clean Conditions}

Table~\ref{tab:averaged_results_clean} presents algorithm performance on clean audio without added noise.

\begin{table}[ht!]
    \centering
    \resizebox{\textwidth}{!}{%
    \begin{tabular}{llccccccc}
        \toprule
        \textbf{Category} & \textbf{Algorithm} & \makecell{\textbf{F1 Score} $\uparrow$} & \makecell{\textbf{CA} $\uparrow$} & \makecell{\textbf{RPA} $\uparrow$} & \makecell{\textbf{RCA} $\uparrow$} & \makecell{\textbf{OA} $\uparrow$} & \makecell{\textbf{GEA} $\uparrow$} & \makecell{\textbf{Harmonic}\\\textbf{Mean} $\uparrow$} \\
        \midrule
        \multirow{4}{*}{\makecell{Neural\\Networks}}
        & PENN \citep{morrison2023cross}            & 92.77 & 91.65  & 85.37 & 86.17 & 86.67 & 87.94 & 89.23 \\
        & CREPE \citep{kim2018crepeconvolutionalrepresentationpitch}     & 79.30 & 88.21  & 70.87 & 71.00 & 92.58 & 80.00 & 80.70 \\
        & BasicPitch \citep{bittner2022lightweightinstrumentagnosticmodelpolyphonic}     & 79.60 & 52.41 & 23.80 & 24.73 & 36.42 & 10.14 & 31.17 \\
        & SwiftF0 (Ours)          & 93.20 & \textbf{94.98} & \textbf{90.07} & 90.13 & \textbf{96.75} & \textbf{97.04} & \textbf{94.07} \\
        \midrule
        \multirow{5}{*}{\makecell{Signal\\Processing}}
        & Praat \citep{BoersmaWeenink2025}            & \textbf{93.63} & 91.92 & 89.20 & \textbf{90.27} & 84.37 & 89.14 & 90.13 \\
        & RAPT \citep{Talkin1995}             & 93.57 & 73.03 & 79.57 & 80.77 & 75.58 & 82.14 & 82.53 \\
        & SWIPE \citep{Camacho2007}            & 89.67 & 84.60  & 85.87 & 88.20 & 63.76 & 78.37 & 81.33 \\
        & YAAPT \citep{Zahorian2008}            & 93.20 & 83.15  & 82.73 & 86.30 & 59.86 & 74.56 & 78.80 \\
        & pYIN \citep{Mauch2014}            & 85.40 & 88.83  & 87.53 & 88.03 & 77.88 & 84.50 & 85.27 \\
        \bottomrule
    \end{tabular}%
    }
    \caption{Averaged performance under clean conditions (no added noise). Compared to the noisy setting (Table~\ref{tab:averaged_results_noise}), most algorithms show marked improvement. SwiftF0 remains top-performing overall.}
    \label{tab:averaged_results_clean}
\end{table}

Under clean conditions, the algorithms' performance shifts considerably, though SwiftF0 maintains superior metrics. SwiftF0 achieves the highest cents accuracy (94.98\%) and highest gross error accuracy (97.04\%), along with the highest harmonic mean (94.07\%). However, Praat also performs strongly in certain areas. It achieves the highest F1 score (93.63\%) and raw chroma accuracy (90.27\%). Crucially, it is important to note that Praat makes more octave errors, and these specific errors are entirely disregarded by RCA. As a result, the RCA score is being "boosted" by correctly identifying the note class even when the pitch is in the wrong octave. For instance, if the true pitch is C4 and the algorithm predicts C5, RCA registers a "hit" (correct pitch class "C"), even though it's an octave off and a clear error for other accuracy measures like RPA.

The results reveal interesting algorithmic characteristics across different noise conditions. PENN shows dramatic improvement when noise is removed, with its harmonic mean increasing from 59.87\% to 89.23\%. Conversely, SwiftF0 demonstrates remarkable stability, with its harmonic mean performance showing only a modest 2.27 percentage point reduction when moving from clean to noisy conditions (from 94.07\% clean to 91.80\% noisy).

Neural methods excel in precision-oriented tasks with superior cents error performance, while traditional methods like YAAPT and Praat show strength in voiced/unvoiced classification.

\subsection{Runtime Performance}

Beyond pitch accuracy, the computational efficiency of pitch detection algorithms is a critical factor, especially for real-time applications, embedded systems, or large-scale data processing. In this section, we present the runtime performance of various pitch detection algorithms when executed on a CPU. Performance is measured both as a relative speed-up factor compared to the slowest algorithm (CREPE) and as absolute execution time in milliseconds (ms) for a 5-second audio file. The benchmarks were conducted on a standard desktop computer with a modern CPU; while absolute execution times may vary across systems, the relative speed-up factors provide a consistent comparison.

\begin{table}[h!]
    \centering
    \begin{tabular}{llcc}
        \toprule
        \textbf{Category} & \textbf{Algorithm} & \textbf{Speed-up ($\times$)} $\uparrow$ & \textbf{Time (ms)} $\downarrow$ \\
        \midrule
        \multirow{4}{*}{\makecell{Neural Network}}
        & PENN \citep{morrison2023cross}          & 6.12  & 919.0   \\
        & CREPE \citep{kim2018crepeconvolutionalrepresentationpitch}    & 1.00  & 5508.3  \\
        & BasicPitch \citep{bittner2022lightweightinstrumentagnosticmodelpolyphonic}    & 20.07 & 280.3   \\
        & SwiftF0 (Ours)       & \textbf{42.42} & \textbf{132.6}  \\
        \midrule
        \multirow{5}{*}{\makecell{Signal Processing}}
        & Praat \citep{BoersmaWeenink2025}         & \textbf{808.99} & \textbf{7.0} \\
        & RAPT \citep{Talkin1995}          & 414.66 & 13.6  \\
        & SWIPE \citep{Camacho2007}         & 42.21  & 133.3  \\
        & YAAPT \citep{Zahorian2008}         & 17.66  & 318.5  \\
        & pYIN \citep{Mauch2014}          & 3.96   & 1420.6 \\
        \bottomrule
    \end{tabular}
    \caption{CPU Runtime Performance. The table displays the average processing time (in milliseconds) and the speed-up factor ($\times$) for each algorithm, relative to CREPE. Lower processing time and higher speed-up indicate better performance.}
    \label{tab:cpu_runtime}
\end{table}

The runtime performance analysis, as shown in Table~\ref{tab:cpu_runtime}, reveals significant differences in computational efficiency among the algorithms. Signal Processing-based algorithms generally outperform Neural Network-based methods in terms of execution speed. Notably, Praat achieves the best performance overall, with an execution time of just 7.0~ms for a 5-second audio file, corresponding to a speed-up factor of 808.99$\times$ relative to CREPE. Among the Neural Network algorithms, SwiftF0 is the fastest, with a speed-up of 42.42$\times$. These results underscore the trade-offs between the typically higher accuracy of Neural Network methods and the superior speed of traditional Signal Processing techniques.

\subsection{Qualitative Analysis}

In addition to our quantitative metrics, we perform a qualitative study on the LJ Speech corpus \citep{ljspeech17} using phoneme-level annotations \citep{https://doi.org/10.5281/zenodo.7499098}. Although the dataset lacks ground-truth fundamental frequency ($F_0$) measurements, we can still judge whether each method’s estimated pitch contour is plausible.

\begin{figure}[htbp]
  \centering
  \includegraphics[width=\textwidth]{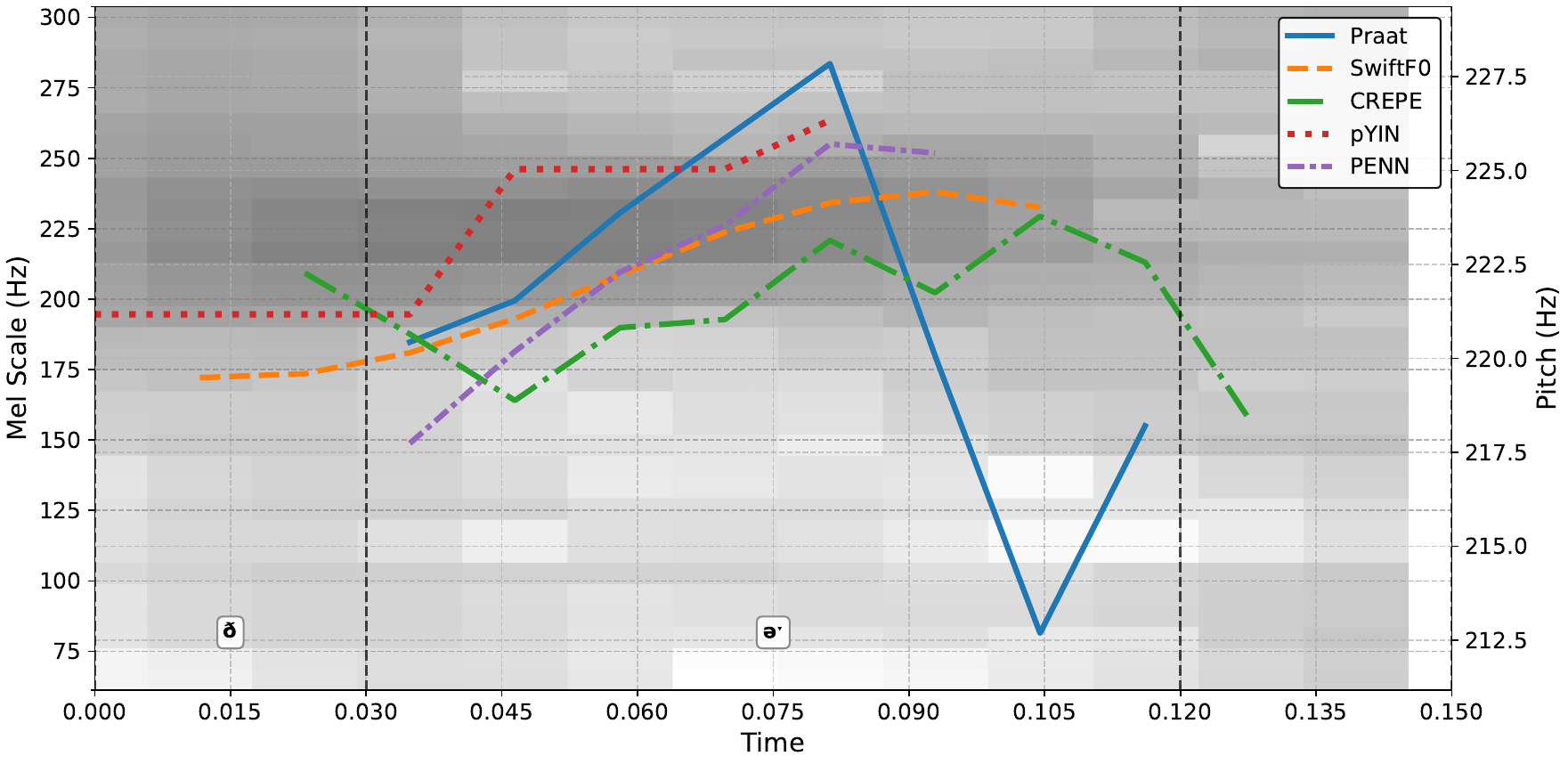}
  \caption{Mel spectrogram and $F_0$ contours for the word “the” (\textipa{/D@/}). SwiftF0 yields a smooth, continuous trajectory, while other algorithms show abrupt jumps or jagged fluctuations.}
  \label{fig:pitch_analysis_the}
\end{figure}

Figure~\ref{fig:pitch_analysis_the} shows the mel spectrogram and $F_0$ traces for the word “the” (\textipa{/D@/}). 

All algorithms place the average pitch between about 212\,Hz and 228\,Hz, but their contour shapes differ markedly. Praat’s trace contains rapid rises and falls that are atypical for natural, continuous speech F0 trajectories and likely represent estimation errors or noise sensitivity. CREPE and pYIN generate jagged lines with frequent small oscillations. PENN is smoother yet still exhibits minor discontinuities. Only SwiftF0 delivers a gently varying, continuous curve that matches the gradual pitch transitions expected in human speech.

\begin{figure}[htbp]
  \centering
  \includegraphics[width=\textwidth]{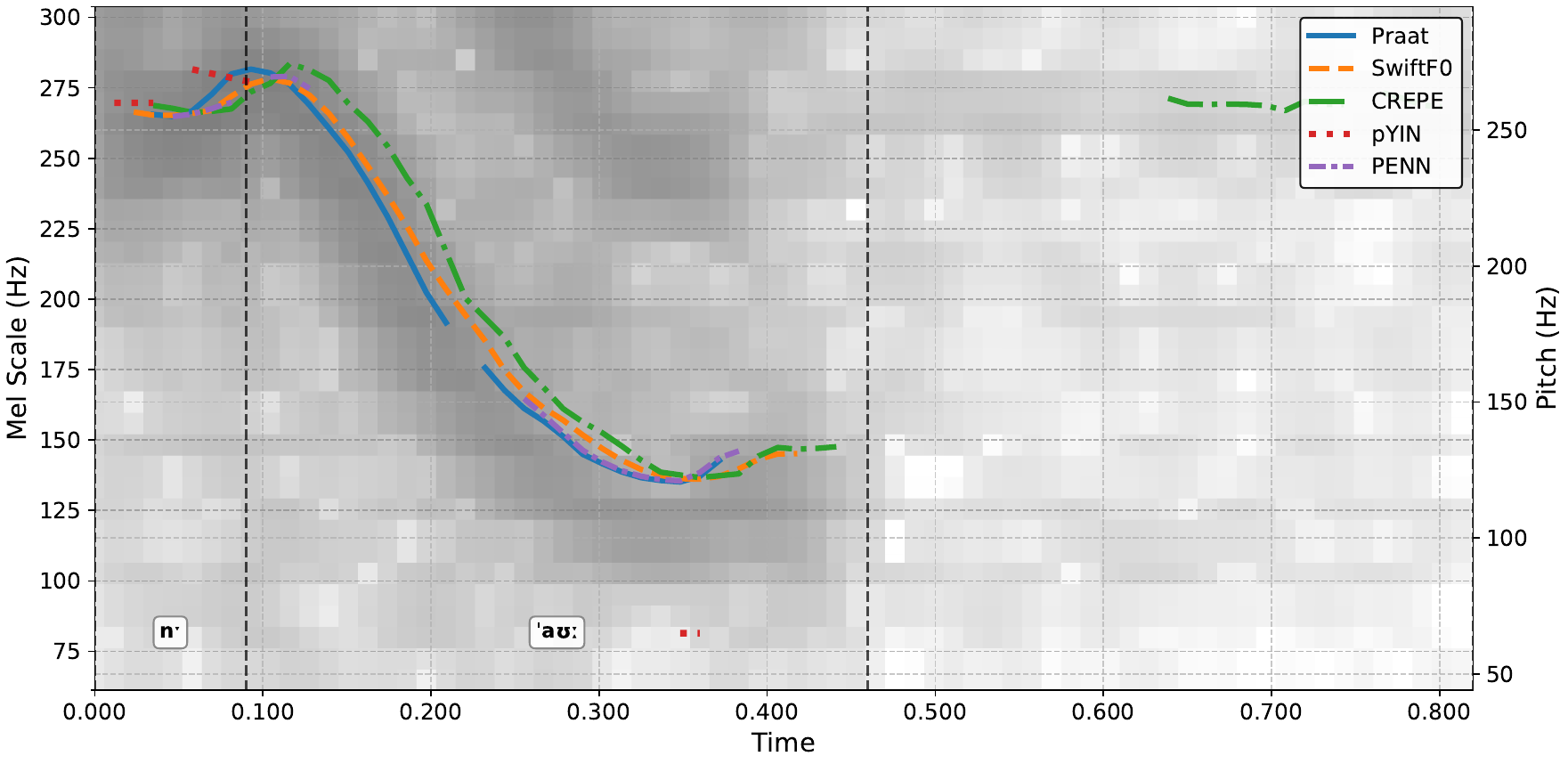}
  \caption{Mel spectrogram and $F_0$ contours for the word “now” (\textipa{/nAU:/}). SwiftF0 maintains a continuous pitch track, whereas other methods introduce gaps or false detections.}
  \label{fig:pitch_analysis_now}
\end{figure}

Figure~\ref{fig:pitch_analysis_now} presents the word “now” (\textipa{/nAU:/}), which includes a longer vowel.

pYIN and PENN have large gaps in the voiced regions, while CREPE reports pitch during unvoiced or silent portions, introducing false detections. In contrast, SwiftF0 produces a fully continuous contour. Praat has a small gap but otherwise follows the falling pitch of the vowel.

These examples demonstrate that a robust $F_0$ estimator must do more than hit the correct frequencies. It must produce smooth transitions and maintain continuity throughout voiced segments. Abrupt spikes, jagged fluctuations, or missing intervals indicate artifacts rather than genuine speech dynamics.

\section{Discussion and Conclusion}

This work introduces three key contributions to monophonic pitch detection.

First, we present SwiftF0, a fast and accurate pitch detection algorithm that outperforms existing approaches in both accuracy and robustness. Under 10\,dB SNR conditions, SwiftF0 achieves a harmonic mean of 91.80\%, surpassing the next-best model by 12\%. Its performance drops by only 2.27 percentage points (from 94.07\% to 91.80\%) in noisy conditions, compared to PENN's sharper 29.36-point decline (from 89.23\% to 59.87\%). SwiftF0 also produces smooth pitch trajectories without requiring median filtering or other special techniques. The only post-processing step is a lightweight local expected value computation, which is directly supported by the model's output distribution. Despite being slower than Praat (132.6\,ms vs. 7.0\,ms for 5 seconds), SwiftF0 is 42.4$\times$ faster than CREPE and more accurate than traditional methods like pYIN. Its efficiency stems from three main factors: a compact architecture with only 95{,}842 parameters, a frequency band selection strategy that reduces input size by 74\%, and the elimination of heavy post-processing.

Second, we introduce SpeechSynth, a synthetic dataset with perfectly accurate pitch annotations generated via controllable speech synthesis. Unlike datasets based on algorithmic extraction or manual correction, SpeechSynth offers an unlimited supply of clean, precisely labeled training data. This provides a powerful resource for speech-focused supervised learning.

Third, we propose a harmonic mean metric that combines six complementary evaluation measures (RPA, Cents Accuracy, Precision, Recall, Octave Accuracy, and Gross Error Accuracy). This metric addresses the shortcomings of individual metrics, discouraging models from excelling at only one aspect while ignoring others.

A key insight from our study is the importance of data diversity. While synthetic datasets like NSynth, MDB-STEM-Synth, and SpeechSynth provide perfect ground truth, training exclusively on them yields suboptimal results. We found that incorporating datasets with algorithmically-derived labels improves generalization. This suggests that such datasets contain real-world acoustic phenomena that synthetic data fails to capture. Although data augmentation with background noise helps bridge this gap, further work is needed to improve the realism of synthetic sources or determine optimal data mixtures.

Another important observation concerns confidence estimation. Our current approach infers confidence from the concentration of probability mass near the predicted pitch. However, since we train only on voiced frames, the resulting scores primarily reflect certainty in pitch class, not voicing decisions. We experimented with adding a dedicated voicing output, but this reduced performance. When the model was allowed to express high voicing confidence despite low pitch certainty, it produced more incorrect pitch predictions, lowering RPA. Future work should explore confidence calibration techniques, entropy-based measures (as used in PENN), or training with unvoiced frames by randomly assigning them pitch bins.

Finally, although SwiftF0 excels on standard metrics, certain application‑specific needs may not be fully captured by these evaluations. We therefore urge future work to assess pitch detectors in the context of representative downstream tasks (e.g., speech synthesis, music transcription), enabling any hidden shortcomings to surface and guiding the development of more comprehensive benchmarks.

\bibliographystyle{unsrtnat}
\bibliography{references}  

\begin{thebibliography}{27}
\providecommand{\natexlab}[1]{#1}
\providecommand{\url}[1]{\texttt{#1}}
\expandafter\ifx\csname urlstyle\endcsname\relax
  \providecommand{\doi}[1]{doi: #1}\else
  \providecommand{\doi}{doi: \begingroup \urlstyle{rm}\Url}\fi

\bibitem[Kim et~al.(2018)Kim, Salamon, Li, and Bello]{kim2018crepeconvolutionalrepresentationpitch}
Jong~Wook Kim, Justin Salamon, Peter Li, and Juan~Pablo Bello.
\newblock {CREPE}: A convolutional representation for pitch estimation, 2018.
\newblock URL \url{https://arxiv.org/abs/1802.06182}.

\bibitem[Boersma and Weenink(2025)]{BoersmaWeenink2025}
Paul Boersma and David Weenink.
\newblock \emph{{Praat}: doing phonetics by computer [Computer program]}, 2025.
\newblock URL \url{https://praat.org}.
\newblock Version 6.4.39, retrieved 13 July 2025 from \url{https://praat.org}.

\bibitem[de~Cheveign{\'e} and Kawahara(2002)]{deCheveigne2002yin}
Alain de~Cheveign{\'e} and Hideki Kawahara.
\newblock {YIN}, a fundamental frequency estimator for speech and music.
\newblock \emph{The Journal of the Acoustical Society of America}, 111\penalty0 (4):\penalty0 1917--1930, 2002.
\newblock \doi{10.1121/1.1458024}.

\bibitem[Mauch and Dixon(2014)]{Mauch2014}
Matthias Mauch and Simon Dixon.
\newblock {pYIN}: A fundamental frequency estimator using probabilistic threshold distributions.
\newblock In \emph{Proceedings of the {IEEE} International Conference on Acoustics, Speech and Signal Processing ({ICASSP})}, pages 659--663, 2014.
\newblock \doi{10.1109/ICASSP.2014.6853681}.
\newblock URL \url{https://ieeexplore.ieee.org/document/6853681}.

\bibitem[Talkin(1995)]{Talkin1995}
David Talkin.
\newblock A robust algorithm for pitch tracking ({RAPT}).
\newblock In W.~B. Kleijn and K.~K. Paliwal, editors, \emph{Speech Coding and Synthesis}, pages 495--518. Elsevier, Amsterdam, 1995.

\bibitem[Camacho(2007)]{Camacho2007}
Andrew Camacho.
\newblock \emph{{SWIPE}: A Sawtooth Waveform Inspired Pitch Estimator for Speech and Music}.
\newblock {Ph.D.} dissertation, University of Florida, 2007.
\newblock URL \url{https://ufdc.ufl.edu/UFE0010168}.

\bibitem[Zahorian and Hu(2008)]{Zahorian2008}
S.~Ali Zahorian and Hui Hu.
\newblock A spectral/temporal method for robust fundamental frequency tracking.
\newblock \emph{The Journal of the Acoustical Society of America}, 123\penalty0 (6):\penalty0 4559--4571, 2008.
\newblock \doi{10.1121/1.2916590}.
\newblock URL \url{https://asa.scitation.org/doi/10.1121/1.2916590}.

\bibitem[Morrison et~al.(2023)Morrison, Hsieh, Pruyne, and Pardo]{morrison2023cross}
Max Morrison, Caedon Hsieh, Nathan Pruyne, and Bryan Pardo.
\newblock Cross-domain neural pitch and periodicity estimation, 2023.

\bibitem[Bittner et~al.(2022)Bittner, Bosch, Rubinstein, Meseguer-Brocal, and Ewert]{bittner2022lightweightinstrumentagnosticmodelpolyphonic}
Rachel~M. Bittner, Juan~Jos{\'e} Bosch, David Rubinstein, Gabriel Meseguer-Brocal, and Sebastian Ewert.
\newblock A lightweight instrument-agnostic model for polyphonic note transcription and multipitch estimation, 2022.
\newblock URL \url{https://arxiv.org/abs/2203.09893}.

\bibitem[Dieleman and Schrauwen(2014)]{6854950}
Sander Dieleman and Benjamin Schrauwen.
\newblock End-to-end learning for music audio.
\newblock In \emph{2014 {IEEE} International Conference on Acoustics, Speech and Signal Processing ({ICASSP})}, pages 6964--6968, 2014.
\newblock \doi{10.1109/ICASSP.2014.6854950}.

\bibitem[Ren et~al.(2022)Ren, Hu, Tan, Qin, Zhao, Zhao, and Liu]{ren2022fastspeech2fasthighquality}
Yi~Ren, Chenxu Hu, Xu~Tan, Tao Qin, Sheng Zhao, Zhou Zhao, and Tie-Yan Liu.
\newblock {FastSpeech 2}: Fast and high-quality end-to-end text to speech, 2022.
\newblock URL \url{https://arxiv.org/abs/2006.04558}.

\bibitem[Luo et~al.(2021)Luo, Tan, Wang, Qin, Li, Zhao, Chen, and Liu]{luo2021lightspeechlightweightfasttext}
Renqian Luo, Xu~Tan, Rui Wang, Tao Qin, Jinzhu Li, Sheng Zhao, Enhong Chen, and Tie-Yan Liu.
\newblock {LightSpeech}: Lightweight and fast text to speech with neural architecture search, 2021.
\newblock URL \url{https://arxiv.org/abs/2102.04040}.

\bibitem[Engel et~al.(2017)Engel, Resnick, Roberts, Dieleman, Eck, Simonyan, and Norouzi]{nsynth2017}
Jesse Engel, Cinjon Resnick, Adam Roberts, Sander Dieleman, Douglas Eck, Karen Simonyan, and Mohammad Norouzi.
\newblock Neural audio synthesis of musical notes with {WaveNet} autoencoders, 2017.

\bibitem[Pirker et~al.(2011)Pirker, Wohlmayr, Petrik, and Pernkopf]{556d2ea3b9c94141a6e0e785972cde62}
Gregor Pirker, Michael Wohlmayr, Stefan Petrik, and Franz Pernkopf.
\newblock A pitch tracking corpus with evaluation on multipitch tracking scenario.
\newblock In \emph{Interspeech - International Conference on Spoken Language Processing}, pages 1509--1512, 2011.

\bibitem[Hsu and Jang(2010)]{5153305}
Chao-Ling Hsu and Jyh-Shing~Roger Jang.
\newblock On the improvement of singing voice separation for monaural recordings using the {MIR-1K} dataset.
\newblock \emph{IEEE Transactions on Audio, Speech, and Language Processing}, 18\penalty0 (2):\penalty0 310--319, 2010.
\newblock \doi{10.1109/TASL.2009.2026503}.

\bibitem[Salamon et~al.(2017)Salamon, Bittner, Bonada, Bosch, G{\'o}mez, and Bello]{salamon2017f0}
Justin Salamon, Rachel~M. Bittner, Jordi Bonada, Juan~J. Bosch, Emilia G{\'o}mez, and Juan~P. Bello.
\newblock An analysis/synthesis framework for automatic {F0} annotation of multitrack datasets.
\newblock In \emph{Proceedings of the 18th International Society for Music Information Retrieval Conference ({ISMIR})}, Suzhou, China, October 2017.

\bibitem[Bittner et~al.(2021)Bittner, Pasalo, Bosch, Meseguer-Brocal, and Rubinstein]{bittner2021vocaditodatasetsolovocals}
Rachel~M. Bittner, Katherine Pasalo, Juan~Jos{\'e} Bosch, Gabriel Meseguer-Brocal, and David Rubinstein.
\newblock {vocadito}: A dataset of solo vocals with $f_0$, note, and lyric annotations, 2021.
\newblock URL \url{https://arxiv.org/abs/2110.05580}.

\bibitem[Shi et~al.(2021)Shi, Bu, Xu, Zhang, and Li]{shi2021aishell3multispeakermandarintts}
Yao Shi, Hui Bu, Xin Xu, Shaoji Zhang, and Ming Li.
\newblock {AISHELL-3}: A multi-speaker mandarin {TTS} corpus and the baselines, 2021.
\newblock URL \url{https://arxiv.org/abs/2010.11567}.

\bibitem[bia(2017)]{biaobei_dataset_2017}
{Biaobei} dataset.
\newblock \url{https://www.data-baker.com/datasets/freeDatasets}, 2017.
\newblock DataBaker (Beijing) Technology Co., Ltd.

\bibitem[Foster et~al.(2015)Foster, Sigtia, Krstulovic, Barker, and Plumbley]{7336899}
Peter Foster, Siddharth Sigtia, Sacha Krstulovic, Jon Barker, and Mark~D. Plumbley.
\newblock {Chime-home}: A dataset for sound source recognition in a domestic environment.
\newblock In \emph{2015 {IEEE} Workshop on Applications of Signal Processing to Audio and Acoustics ({WASPAA})}, pages 1--5, 2015.
\newblock \doi{10.1109/WASPAA.2015.7336899}.

\bibitem[Jadoul et~al.(2018)Jadoul, Thompson, and de~Boer]{Jadoul2018}
Yannick Jadoul, Benjamin Thompson, and Bas de~Boer.
\newblock Introducing {Parselmouth}: A {Python} interface to {Praat}.
\newblock \emph{Journal of Phonetics}, 71:\penalty0 1--15, 2018.
\newblock \doi{10.1016/j.wocn.2018.07.001}.

\bibitem[{SPTK developers}(2025)]{SPTK2025}
{SPTK developers}.
\newblock \emph{Speech Signal Processing Toolkit ({SPTK})}, 2025.
\newblock URL \url{https://sp-nitech.github.io/sptk/latest/main/pitch.html}.
\newblock Includes {RAPT} and {SWIPE} pitch extraction algorithms, retrieved 13 July 2025.

\bibitem[McFee et~al.(2023)]{BMcFee2023_librosa}
Brian McFee et~al.
\newblock {librosa}: Audio and music signal analysis in {Python}, 2023.
\newblock URL \url{https://librosa.org/}.

\bibitem[{pYAAPT maintainers}(2023)]{pYAAPT2023}
{pYAAPT maintainers}.
\newblock {pYAAPT}: Python implementation of the {YAAPT} pitch tracker.
\newblock \url{https://bjbschmitt.github.io/AMFM_decompy/}, 2023.

\bibitem[Morrison(2023)]{TorchCREPE2023}
Max Morrison.
\newblock {TorchCREPE}: {PyTorch} implementation of the {CREPE} pitch tracker.
\newblock \url{https://github.com/maxrmorrison/torchcrepe}, 2023.

\bibitem[Ito and Johnson(2017)]{ljspeech17}
Keith Ito and Linda Johnson.
\newblock The {LJ Speech Dataset}.
\newblock \url{https://keithito.com/LJ-Speech-Dataset/}, 2017.

\bibitem[Taubert(2023)]{https://doi.org/10.5281/zenodo.7499098}
Stefan Taubert.
\newblock {LJ Speech} - aligned {IPA} transcriptions, 2023.
\newblock URL \url{https://zenodo.org/doi/10.5281/zenodo.7499098}.

\end{thebibliography}






\end{document}